# A Method for Solving Linearized Vlasov Equation for Low-Frequency Long-Wavelength Electromagnetic Modes in Inhomogeneous Plasmas

Bamandas Basu

Research Affiliate, High Energy Plasma Physics Group, Laboratory for Nuclear Science
MIT, Cambridge, Massachusetts, USA

**Abstract**

A method for solving linearized Vlasov equation for low-frequency, long-wavelength electromagnetic modes in magnetically confined inhomogeneous plasmas is described. The relevant non-local solution that includes the lowest-significant-order effects of inhomogeneities is obtained from the solutions of three simple equations by means of elementary algebra. The method appears to be more convenient than the commonly used method of integration along the unperturbed particle orbits and should be of interest to students of theoretical plasma physics.

**Introduction**

The low-frequency and long-wavelength plasma modes, with frequencies smaller than the ion gyrofrequency and wavelengths longer than the ion gyroradius, are frequently observed in both space and laboratory plasmas. Well-known examples are the various kinds of drift waves (both electrostatic and electromagnetic) that are excited in inhomogeneous plasmas immersed in inhomogeneous magnetic field. Anomalous particle and energy transports enabled by these drift waves, when they are driven unstable, can exceed the corresponding collisional transports by several orders of magnitude. Consequently, these waves are of considerable importance to plasma physicists.

The first and the most important step in the derivation of the dispersion equation for the plasma modes in collisionless inhomogeneous plasmas is the solution of the relevant linearized Vlasov equation. The commonly used method of solution[1] is to integrate the linearized Vlasov equation along the particle orbits in the equilibrium magnetic field. This yields expressions for the perturbed charged particle distribution functions, which are used in the calculations of the induced charged and current densities leading to the derivation of the dispersion equation via Maxwell equations. In practice, however, calculation of the particle orbits alone can be a difficult task depending on the complexity of the equilibrium magnetic field. In fact, even for a rather simple equilibrium magnetic field configuration, such as that represented by $\mathbf{B}_0 = B_0\left(1 + x/L_m\right)\mathbf{e}_z$ in the Cartesian coordinate system, the particle orbit calculation recovering important lowest-order particle drifts is a non-trivial task. Moreover, if the inclusion of the non-local effects in the dispersion equation is necessary, as is generally the case for waves in inhomogeneous plasmas, the algebra involved in the method of integration along the unperturbed particle orbits can be very tedious.

In this short note, we present an alternative method of solving the linearized Vlasov equation that is suitable for studying low-frequency, long-wavelength electromagnetic modes in magnetically confined inhomogeneous plasmas.

**Method of Solution**

We consider the problem of electromagnetic drift waves in a plasma with inhomogeneity only along the $x-$direction such that equilibrium density and temperature are represented by $n_\alpha(x) = n_{\alpha 0}(1 - x/L_{n\alpha})$ and $T_\alpha(x) = T_{\alpha 0}(1 - x/L_{T\alpha})$, respectively, where $\alpha = e, i$ denote the charged particle species (electrons and ions). The equilibrium magnetic field in plane geometry is modelled by

$$\mathbf{B}_0(x) = \hat{B}_0 \left[ \left(1 + \frac{x}{L_m}\right) \mathbf{e}_z + \frac{x}{L_s} \mathbf{e}_y \right], \tag{1}$$

which represents a sheared magnetic field with a gradient resulting from the diamagnetic drift currents due to the plasma inhomogeneities. Here $L_m$ and $L_s$ are the scale length of the magnetic field gradient and the magnetic shear length, respectively. All the inhomogeneities are assumed to be weak, meaning that $x/L_{n\alpha}$, $x/L_{T\alpha}$, $x/L_m$ and $x/L_s$ are all much less than unity. A similar problem was solved by Catto et al.[2] by the method of integration mentioned above.

The distribution function $f_{\alpha 0}$ that describes the considered equilibrium plasma state is

$$f_{\alpha 0} = \left\{ 1 - \left(x + \frac{v_y}{\Omega_\alpha}\right) \left[ \frac{1}{L_{n\alpha}} + \frac{1}{L_{T\alpha}} \left( \frac{m_\alpha v^2}{2T_{\alpha 0}} - \frac{3}{2} \right) \right] \right\} F_{\alpha 0}, \tag{2}$$

where $v^2 \equiv v_x^2 + v_y^2 + v_z^2$,

$$F_{\alpha 0} = n_{\alpha 0} \left( \frac{m_\alpha}{2\pi T_{\alpha 0}} \right)^{3/2} \exp\left( -\frac{m_\alpha v^2}{2T_{\alpha 0}} \right), \tag{3}$$

and $\Omega_\alpha \equiv q_\alpha \hat{B}_0 / (m_\alpha c)$, $q_\alpha$ being the charge of the particle species $\alpha$ including the sign of the charge. The small equilibrium current in the $z$–direction, which is required for self-consistency with the sheared magnetic field, has been omitted for simplicity. We remark here that $f_{\alpha 0}$ is constructed from the constants of motion, namely, the total energy $m_\alpha v^2/2$ and the $y$–component of canonical momentum, which follow from the symmetry arguments without having to solve the particle orbit equations when weak inhomogeneity is considered [see e.g., Ref. 3]. It can indeed be verified that $f_{\alpha 0}$, given by Eqs. (2) and (3), is a solution of the corresponding Vlasov equation for $f_{\alpha 0}$ to the lowest order of $x/L_m \ll 1$.

The linearized Vlasov equation to be solved is

$$\left[\frac{\partial}{\partial t} + \mathbf{v}\cdot\nabla + \frac{q_\alpha}{m_\alpha c}(\mathbf{v}\times\mathbf{B}_0)\cdot\frac{\partial}{\partial \mathbf{v}}\right]\hat{f}_\alpha + \frac{q_\alpha}{m_\alpha}\left(\hat{\mathbf{E}} + \frac{1}{c}\mathbf{v}\times\hat{\mathbf{B}}\right)\cdot\frac{\partial}{\partial \mathbf{v}} f_{\alpha 0} = 0, \quad (4)$$

where $\hat{f}_\alpha$, $\hat{\mathbf{E}}$ and $\hat{\mathbf{B}}$ denote the perturbations of the respective quantities. We assume that all perturbations have space-time dependence of the form: $\hat{A} = \tilde{A}(x)\exp[i(ky - \omega t)]$. The plane $x = 0$ is located such that $\mathbf{k}\cdot\mathbf{B}_0(x=0) = 0$, and the parallel component of the wave vector is defined as $k_\parallel \equiv k(x/L_s)$. Eliminating $\tilde{E}_x$, $\tilde{E}_z$ and $\tilde{B}_y$ in favor of $\tilde{E}_y$, $\tilde{B}_x$ and $\tilde{B}_z$ with the help of

$$\nabla\times\hat{\mathbf{E}} = -\frac{1}{c}\frac{\partial}{\partial t}\hat{\mathbf{B}} \quad (5)$$

and $\nabla\cdot\hat{\mathbf{B}} = 0$, Eq. (4) can be rewritten as

$$\left[\omega - kv_y + iv_x\frac{\partial}{\partial x} + i\frac{q_\alpha}{m_\alpha c}(\mathbf{v}\times\mathbf{B}_0)\cdot\frac{\partial}{\partial \mathbf{v}}\right]\tilde{f}_\alpha$$
$$= \frac{q_\alpha}{kT_{\alpha 0}}\left\{\left[1 - \frac{1}{L'_\alpha}\left(x + \frac{v_y}{\Omega_\alpha}\right)\right]\left[\left(v_x\frac{\partial}{\partial x} + ikv_y\right)\tilde{E}_y + i\frac{\omega}{c}(v_z\tilde{B}_x - v_x\tilde{B}_z)\right]\right. \quad (6)$$
$$\left. + i\frac{kT_{\alpha 0}}{m_\alpha \Omega_\alpha L_\alpha}\left[\tilde{E}_y + \frac{1}{c}(v_z\tilde{B}_x - v_x\tilde{B}_z)\right]\right\}F_{\alpha 0}$$

where we have used the expression for $f_{\alpha 0}$ and introduced the notations

$$\frac{1}{L_\alpha} \equiv \frac{1}{L_{n\alpha}} + \frac{1}{L_{T\alpha}}\left(\frac{m_\alpha v^2}{2T_{\alpha 0}} - \frac{3}{2}\right) \tag{7}$$

and

$$\frac{1}{L'_\alpha} \equiv \frac{1}{L_{n\alpha}} + \frac{1}{L_{T\alpha}}\left(\frac{m_\alpha v^2}{2T_{\alpha 0}} - \frac{5}{2}\right). \tag{8}$$

The first step in the method of solution is to express Eq. (6) in terms of the cylindrical coordinates $(v_\perp, \varphi, v_\parallel)$ in velocity space that are defined in the following way[2]:

$$v_\parallel \equiv \frac{1}{|\mathbf{B}_0|}(\mathbf{v}\cdot\mathbf{B}_0) = \frac{1}{\Delta}\left[\left(1+\frac{x}{L_m}\right)v_z + \frac{x}{L_s}v_y\right], \tag{9}$$

$$v_\perp \cos\varphi = v_x, \tag{10}$$

$$v_\perp \sin\varphi = \frac{1}{\Delta}\left[\left(1+\frac{x}{L_m}\right)v_y + \frac{x}{L_s}v_z\right], \tag{11}$$

where

$$\Delta \equiv \left[\left(1+\frac{x}{L_m}\right)^2 + \frac{x^2}{L_s^2}\right]^{1/2}. \tag{12}$$

It can be verified that $v_\perp^2 + v_\parallel^2 = v_x^2 + v_y^2 + v_z^2 = v^2$ and that $d\mathbf{v} = v_\perp dv_\perp dv_\parallel d\varphi$. For weak inhomogeneities $(x/L_m \ll 1, x/L_s \ll 1)$, we may take $\Delta \simeq 1 + x/L_m$ and then find from Eqs. (9) – (11) that

$$v_x = v_\perp \cos\varphi, \quad v_y = v_\perp \sin\varphi + \frac{x}{L_s}v_\parallel, \quad v_z = v_\parallel - \frac{x}{L_s}v_\perp \sin\varphi \tag{13}$$

and

$$\frac{\partial}{\partial v_x} = \cos\varphi \frac{\partial}{\partial v_\perp} - \frac{\sin\varphi}{v_\perp} \frac{\partial}{\partial \varphi}, \tag{14}$$

$$\frac{\partial}{\partial v_y} = \sin\varphi \frac{\partial}{\partial v_\perp} + \frac{\cos\varphi}{v_\perp} \frac{\partial}{\partial \varphi} + \frac{x}{L_s} \frac{\partial}{\partial v_\parallel}, \tag{15}$$

$$\frac{\partial}{\partial v_z} = \frac{\partial}{\partial v_\parallel} - \frac{x}{L_s}\left(\sin\varphi \frac{\partial}{\partial v_\perp} + \frac{\cos\varphi}{v_\perp} \frac{\partial}{\partial \varphi}\right), \tag{16}$$

when only the terms of order $x/L_s$ are retained. In terms of the cylindrical coordinates, Eq. (6) becomes

$$\left\{\omega - k_\parallel v_\parallel - k v_\perp \sin\varphi + i v_\perp \cos\varphi \frac{\partial}{\partial x} - i\Omega_\alpha\left(1 + \frac{x}{L_m}\right)\frac{\partial}{\partial \varphi}\right\} \tilde{f}_\alpha(x, v_\perp, v_\parallel, \varphi)$$

$$= \frac{q_\alpha}{kT_{\alpha 0}}\left\{\left[1 - \frac{1}{L'_\alpha}\left(x + \frac{v_\perp}{\Omega_\alpha}\sin\varphi\right)\right]\left[\left(v_\perp \cos\varphi \frac{\partial}{\partial x} + ikv_\perp \sin\varphi\right)\tilde{E}_y + i\frac{\omega}{c}\left(v_\parallel \tilde{B}_x - v_\perp \cos\varphi \tilde{B}_z\right)\right] \right. \tag{17}$$

$$\left. + ik_\parallel\left(v_\parallel \tilde{E}_y - \frac{\omega}{ck} v_\perp \sin\varphi \tilde{B}_x\right) + i\frac{kT_{\alpha 0}}{m_\alpha \Omega_\alpha L_\alpha}\left[\tilde{E}_y + \frac{1}{c}\left(v_\parallel \tilde{B}_x - v_\perp \cos\varphi \tilde{B}_z\right)\right]\right\} F_{\alpha 0}.$$

Here $k_\parallel \equiv k(x/L_s)$ and only the leading-order terms of inhomogeneity have been retained.

Now, we use the solution

$$\tilde{f}_\alpha(x, v_\perp, v_\parallel, \varphi) = \sum_{n=-\infty}^{+\infty} \tilde{g}_{\alpha,n}(x, v_\perp, v_\parallel) \exp(in\varphi) \tag{18}$$

in Eq. (17) and equating the coefficients of $\exp(in\varphi)$, for $n = 0, \pm 1, \pm 2$, etc., obtain the following interconnected chain of equations

$$(\omega - k_\parallel v_\parallel)\tilde{g}_{\alpha,0} + i\frac{v_\perp}{2}\left(\frac{\partial}{\partial x} - k\right)\tilde{g}_{\alpha,1} + i\frac{v_\perp}{2}\left(\frac{\partial}{\partial x} + k\right)\tilde{g}_{\alpha,-1}$$

$$= i\frac{q_\alpha}{kT_{\alpha 0}}\left\{\left(k_\parallel v_\parallel + \frac{kT_{\alpha 0}}{m_\alpha \Omega_\alpha L_\alpha} - \frac{kv_\perp^2}{2\Omega_\alpha L'_\alpha}\right)\tilde{E}_y + \left[\omega\left(1 - \frac{x}{L'_\alpha}\right) + \frac{kT_{\alpha 0}}{m_\alpha \Omega_\alpha L_\alpha}\right]\frac{v_\parallel}{c}\tilde{B}_x\right\} F_{\alpha 0}, \tag{19}$$

$$\left[\omega - k_\| v_\| \pm \Omega_\alpha \left(1 + \frac{x}{L_m}\right)\right] \tilde{g}_{\alpha,\pm 1} + i \frac{v_\perp}{2} \left(\frac{\partial}{\partial x} \pm k\right) \tilde{g}_{\alpha,0} + i \frac{v_\perp}{2} \left(\frac{\partial}{\partial x} \mp k\right) \tilde{g}_{\alpha,\pm 2}$$

$$= \frac{q_\alpha}{2kT_{\alpha 0}} v_\perp \left\{\left(1 - \frac{x}{L'_\alpha}\right)\left[\left(\frac{\partial}{\partial x} \pm k\right)\tilde{E}_y - i\frac{\omega}{c}\tilde{B}_z\right] - i\frac{kT_{\alpha 0}}{m_\alpha \Omega_\alpha L_\alpha c}\tilde{B}_z \mp \frac{\omega}{c}\left(\frac{x}{L_s} + \frac{1}{L'_\alpha}\frac{v_\|}{\Omega_\alpha}\right)\tilde{B}_x\right\} F_{\alpha 0}, \quad (20)$$

$$\left[\omega - k_\| v_\| \pm 2\Omega_\alpha \left(1 + \frac{x}{L_m}\right)\right] \tilde{g}_{\alpha,\pm 2} + i \frac{v_\perp}{2} \left(\frac{\partial}{\partial x} \pm k\right) \tilde{g}_{\alpha,\pm 1} + i \frac{v_\perp}{2} \left(\frac{\partial}{\partial x} \mp k\right) \tilde{g}_{\alpha,\pm 3}$$

$$= \pm i \frac{q_\alpha}{4kT_{\alpha 0}} \frac{v_\perp^2}{\Omega_\alpha L'_\alpha} \left[\left(\frac{\partial}{\partial x} \pm k\right)\tilde{E}_y - i\frac{\omega}{c}\tilde{B}_z\right] F_{\alpha 0}, \quad (21)$$

and for all $|n| = p > 2$,

$$\left[\omega - k_\| v_\| \pm p\Omega_\alpha \left(1 + \frac{x}{L_m}\right)\right] \tilde{g}_{\alpha,\pm p} + i \frac{v_\perp}{2} \left(\frac{\partial}{\partial x} \pm k\right) \tilde{g}_{\alpha,\pm(p-1)} + i \frac{v_\perp}{2} \left(\frac{\partial}{\partial x} \mp k\right) \tilde{g}_{\alpha,\pm(p+1)} = 0. \quad (22)$$

Clearly, Eqs. (19) – (22) do not form a closed set, as the equation for $\tilde{g}_{\alpha,\pm p}$ involves $\tilde{g}_{\alpha,\pm(p+1)}$ where $p = 0, 1, 2, \ldots$. However, it may also be observed from Eqs. (20) – (22) that

$$\tilde{g}_{\alpha,\pm p} \sim \frac{v_\perp}{p\Omega_\alpha}\left(k, \frac{\partial}{\partial x}\right) \tilde{g}_{\alpha,\pm(p-1)} \quad (23)$$

for $p = 0, 1, 2, \ldots$. Therefore, for the description of long – wavelength and localized modes with the distance of localization larger than the gyroradii of the charged particles, such that $(kv_\perp/\Omega_\alpha) < 1$ and $(v_\perp/\Omega_\alpha)(\partial/\partial x) < 1$, the chain of equations may be truncated by choosing $\tilde{g}_{\alpha,\pm p} = 0$ for all $p \geq l$, where the value of $l$ is determined by the desired level of accuracy of the description. In particular, we note that, according to Eq. (18),

$$\tilde{f}_\alpha(x, v_\perp, v_\|, \varphi) = \tilde{g}_{\alpha,0} + (\tilde{g}_{\alpha,1} + \tilde{g}_{\alpha,-1})\cos\varphi + i(\tilde{g}_{\alpha,1} - \tilde{g}_{\alpha,-1})\sin\varphi \quad (24)$$

is sufficient for the calculation of the perturbed charge and current densities and, hence, for the derivation of the dispersion equation. Then, since we require an expression for $\tilde{f}_\alpha$ with terms up to orders $\left(k^2 v_\perp^2 / \Omega_\alpha^2\right)$ and $\left(v_\perp^2 / \Omega_\alpha^2\right)\left(\partial^2 / \partial x^2\right)$ for the description of the non-local behavior of the modes to lowest-significant-order, we may choose $\tilde{g}_{\alpha,\pm p} = 0$ for all $p \geq 2$. That is, we need to solve the three equations given by Eq. (19) and Eqs. (20) with $\tilde{g}_{\alpha,\pm 2} = 0$.

In solving the equations, we use $|x/L| \ll 1$, $|v_\parallel / \Omega_\alpha| \ll 1$, $|v_\perp / \Omega_\alpha| \ll 1$ and $|\bar{\omega}/\Omega_\alpha| \ll 1$, where $L$ denotes the scale length for all inhomogeneities and $\bar{\omega} \equiv \omega - k_\parallel v_\parallel$. We also use $k \sim (\partial/\partial x) > 1/L$ and $1/L \sim (\bar{\omega}/\Omega_\alpha)(\partial/\partial x) \sim (\bar{\omega}/\Omega_\alpha)k$. Substituting $\tilde{g}_{\alpha,\pm 1}$, obtained from Eqs. (20), into Eq. (19) and retaining terms up to order $k^2 v_\perp^2 / \Omega_\alpha^2 \sim \left(v_\perp^2/\Omega_\alpha^2\right)\left(\partial^2/\partial x^2\right)$ [this is achieved by retaining corrections of the order of $(\bar{\omega}/\Omega_\alpha)$ in $\tilde{g}_{\alpha,\pm 1}$], we find

$$\left[1 - \frac{\omega_{D\alpha}}{\bar{\omega}} - \frac{v_\perp^2}{2\Omega_\alpha^2}\left(\frac{\partial^2}{\partial x^2} - k^2\right)\right]\tilde{g}_{\alpha,0}$$
$$= i\frac{q_\alpha}{kT_{\alpha 0}} F_{\alpha 0}\left\{\frac{1}{\bar{\omega}}\left[\left(k_\parallel v_\parallel + \omega_{D\alpha}\right)\tilde{E}_y + \frac{\omega}{c}\left(v_\parallel \tilde{B}_x - i\frac{kv_\perp^2}{2\Omega_\alpha}\tilde{B}_z\right)\right]\right. \quad (25)$$
$$\left. - \frac{\omega_{*n\alpha}}{\bar{\omega}}\left[\tilde{E}_y + \frac{1}{c}\left(v_\parallel \tilde{B}_x - i\frac{kv_\perp^2}{2\Omega_\alpha}\tilde{B}_z\right)\right]\left[1 + \eta_\alpha\left(\frac{m_\alpha v^2}{2T_{\alpha 0}} - \frac{3}{2}\right)\right] + \frac{v_\perp^2}{2\Omega_\alpha^2}\left(\frac{\partial^2}{\partial x^2} - k^2\right)\tilde{E}_y\right\}$$

Here $\omega_{D\alpha} \equiv kv_\perp^2 / (2\Omega_\alpha L_m)$ is the frequency associated with the particle drift in the equilibrium magnetic field with a gradient,

$$\omega_{*n\alpha} \equiv -k\frac{T_{\alpha 0}}{m_\alpha \Omega_\alpha L_{n\alpha}} \quad \text{and} \quad \eta_\alpha \equiv \frac{L_{n\alpha}}{L_{T\alpha}}. \quad (26)$$

Since $\omega_{D\alpha}/\bar{\omega} \sim \left[v_\perp^2/(2\Omega_\alpha^2)\right]\left(\partial^2/\partial x^2\right) \sim k^2 v_\perp^2 / (2\Omega_\alpha^2) < 1$, Eq. (25) can be solved by an iteration resulting in

$$\tilde{g}_{\alpha,0} = -i\frac{q_\alpha}{kT_{\alpha 0}} F_{\alpha 0} \left\{ \tilde{E}_y - \frac{\omega}{\bar{\omega}}\left[1 - \frac{\omega_{*n\alpha}}{\omega}\left(1 + \eta_\alpha\left(\frac{m_\alpha v^2}{2T_{\alpha 0}} - \frac{3}{2}\right)\right)\right] \right. \\ \left. \times \left[\left(1 + \frac{\omega_{D\alpha}}{\bar{\omega}} + \frac{v_\perp^2}{2\Omega_\alpha^2}\left(\frac{\partial^2}{\partial x^2} - k^2\right)\right)\left(\tilde{E}_y + \frac{v_\parallel}{c}\tilde{B}_x - i\frac{k v_\perp^2}{2c\Omega_\alpha}\tilde{B}_z\right)\right] \right\} \quad (27)$$

where only the leading order terms have been retained. Once $\tilde{g}_{\alpha,0}$ has been found, it is then a simple matter to find $\tilde{g}_{\alpha,\pm 1}$ from Eq. (20) with $\tilde{g}_{\alpha,\pm 2} = 0$. Keeping terms up to order $k^2 v_\perp^2 / (2\Omega_\alpha^2)$ we find

$$\tilde{g}_{\alpha,1} + \tilde{g}_{\alpha,-1} \simeq \frac{q_\alpha}{kT_{\alpha 0}} \frac{\omega}{\bar{\omega}} \left\{ 1 - \frac{\omega_{*n\alpha}}{\bar{\omega}}\left[1 + \eta_\alpha\left(\frac{m_\alpha v^2}{2T_{\alpha 0}} - \frac{3}{2}\right)\right] \right\} \\ \times \left\{ \frac{k v_\perp}{\Omega_\alpha}\left(\tilde{E}_y + \frac{v_\parallel}{c}\tilde{B}_x\right) - i\left(\frac{k^2 v_\perp^2}{2\Omega_\alpha^2} - \frac{\bar{\omega}^2}{\Omega_\alpha^2}\right)\frac{v_\perp}{c}\tilde{B}_z \right\} F_{\alpha 0} \quad (28)$$

and

$$\tilde{g}_{\alpha,1} - \tilde{g}_{\alpha,-1} \simeq \frac{q_\alpha}{kT_{\alpha 0}} \frac{\omega}{\bar{\omega}} \left\{ 1 - \frac{\omega_{*n\alpha}}{\bar{\omega}}\left[1 + \eta_\alpha\left(\frac{m_\alpha v^2}{2T_{\alpha 0}} - \frac{3}{2}\right)\right] \right\} \\ \times \left[ \frac{v_\perp}{\Omega_\alpha}\frac{\partial}{\partial x}\left(\tilde{E}_y + \frac{v_\parallel}{c}\tilde{B}_x\right) - i\frac{v_\perp}{c}\left(\frac{k v_\perp^2}{2\Omega_\alpha^2}\frac{\partial}{\partial x} + \frac{\bar{\omega}}{\Omega_\alpha}\right)\tilde{B}_z \right] F_{\alpha 0} \quad (29)$$

Substitution of Eqs. (27) – (29) into Eq. (24) yields the approximate solution of the linearized Vlasov equation including the essential non-local effects. It is sufficient for the derivation of the electromagnetic dispersion relation as it determines the required perturbed charge and current densities. In fact, the perturbed charge density is given by

$$\tilde{\rho}_\alpha \equiv q_\alpha \int d\mathbf{v}\, \tilde{f}_\alpha = 2\pi q_\alpha \int dv_\perp dv_\parallel v_\perp \tilde{g}_{\alpha,0}, \quad (30)$$

while the $x-$ and $z-$ components of the perturbed current density are given by

$$\tilde{j}_x \equiv \sum_\alpha q_\alpha \int d\mathbf{v}\, v_x \tilde{f}_\alpha = \pi \sum_\alpha q_\alpha \int dv_\perp dv_\parallel v_\perp^2 \left(\tilde{g}_{\alpha,1} + \tilde{g}_{\alpha,-1}\right), \quad (31)$$

and

$$\tilde{j}_z \equiv \sum_\alpha q_\alpha \int d\mathbf{v}\, v_z \tilde{f}_\alpha = 2\pi \sum_\alpha q_\alpha \int dv_\perp dv_\parallel v_\perp \left[ v_\parallel \tilde{g}_{\alpha,0} - i\frac{x}{2L_s} v_\perp \left( \tilde{g}_{\alpha,1} - \tilde{g}_{\alpha,-1} \right) \right], \quad (32)$$

respectively. The contribution to $\tilde{j}_z$ of the second term within the parentheses, when compared with that of the first term, is negligible.

If higher-order corrections to $\tilde{g}_{\alpha,0}$ and $\left( \tilde{g}_{\alpha,1} \pm \tilde{g}_{\alpha,-1} \right)$ are needed in a particular problem of interest, they can be obtained by including the equations for $\tilde{g}_{\alpha,\pm 2}$ with $\tilde{g}_{\alpha,\pm 3} = 0$ and so on in the above analysis.

As a check, let us derive the electrostatic dispersion relation that follows from the solution presented above. For this we take $\tilde{E}_y = -ik\tilde{\phi}$ and $\tilde{B}_x = \tilde{B}_y = \tilde{B}_z = 0$. Then, $\tilde{g}_{\alpha,0}$ is given by

$$\begin{aligned}
\tilde{g}_{\alpha,0} = -\frac{q_\alpha}{T_{\alpha 0}} F_{\alpha 0} &\left\{ 1 - \frac{\omega}{\bar{\omega}} \left[ 1 - \frac{\omega_{*n\alpha}}{\omega} \left( 1 + \eta_\alpha \left( \frac{m_\alpha v^2}{2T_{\alpha 0}} - \frac{3}{2} \right) \right) \right] \right. \\
&\left. \times \left[ 1 + \frac{\omega_{D\alpha}}{\bar{\omega}} + \frac{v_\perp^2}{2\Omega_\alpha^2} \left( \frac{\partial^2}{\partial x^2} - k^2 \right) \right] \right\} \tilde{\phi}
\end{aligned} \quad (33)$$

and the Poisson equation, $k^2 \tilde{\phi} = 4\pi \sum_\alpha \tilde{\rho}_\alpha$, yields the non-local dispersion relation

$$A \left( \frac{\partial^2}{\partial x^2} - k^2 \right) \tilde{\phi} - B\tilde{\phi} \simeq 0, \quad (34)$$

where

$$A \equiv \sum_\alpha \frac{2\pi^2 q_\alpha^2}{k^2 \Omega_\alpha^2 T_{\alpha 0}} \int dv_\perp dv_\parallel v_\perp^3 \frac{\omega}{\bar{\omega}} \left\{ 1 - \frac{\omega_{*n\alpha}}{\omega} \left[ 1 + \eta_\alpha \left( \frac{m_\alpha v^2}{2T_{\alpha 0}} - \frac{3}{2} \right) \right] \right\} F_{\alpha 0} \quad (35)$$

and

$$B \equiv 1 + \sum_\alpha \frac{8\pi^2 q_\alpha^2}{k^2 T_{\alpha 0}} \int dv_\perp dv_\parallel v_\perp \left\{ 1 - \frac{\omega}{\bar{\omega}} \left( 1 + \frac{\omega_{D\alpha}}{\bar{\omega}} \right) \left[ 1 - \frac{\omega_{*n\alpha}}{\omega} \left( 1 + \eta_\alpha \left( \frac{m_\alpha v^2}{2T_{\alpha 0}} - \frac{3}{2} \right) \right) \right] \right\} F_{\alpha 0}. \quad (36)$$

Using $F_{\alpha 0}$, given by Eq. (3) and carrying out the velocity integrations, we find

$$A = \frac{1}{2}\sum_\alpha \frac{\omega_{p\alpha}^2 m_\alpha}{k^2 \Omega_\alpha^2}\left\{\left[1-\frac{\omega_{*n\alpha}}{\omega}\left(1+\frac{1}{2}\eta_\alpha\right)\right]\left[1+W(\varsigma_\alpha)\right]-\eta_\alpha\frac{\omega_{*n\alpha}}{\omega}\varsigma_\alpha^2 W(\varsigma_\alpha)\right\} \qquad (37)$$

and

$$\begin{aligned}B \equiv 1 &+ \sum_\alpha \frac{\omega_{p\alpha}^2 m_\alpha}{k^2 T_{\alpha 0}}\left\{1-\left[1-\frac{\omega_{*n\alpha}}{\omega}\left(1-\frac{1}{2}\eta_\alpha\right)\right]\left[1+W(\varsigma_\alpha)\right]+\eta_\alpha\frac{\omega_{*n\alpha}}{\omega}\varsigma_\alpha^2 W(\varsigma_\alpha)\right\} \\ &- 2\sum_\alpha \frac{\omega_{p\alpha}^2 m_\alpha}{k^2 T_{\alpha 0}}\left(\frac{\omega\bar{\omega}_{D\alpha} m_\alpha}{k_\parallel^2 T_{\alpha 0}}\right)\left\{\left[1-\frac{\omega_{*n\alpha}}{\omega}\left(1-\frac{1}{2}\eta_\alpha\right)\right]W(\varsigma_\alpha)+\frac{1}{2}\eta_\alpha\frac{\omega_{*n\alpha}}{\omega}\left[1-2\varsigma_\alpha^2 W(\varsigma_\alpha)\right]\right\}\end{aligned} \qquad (38)$$

where we have defined $\bar{\omega}_{D\alpha} \equiv kT_{\alpha 0}/(2m_\alpha \Omega_\alpha L_m)$, $\varsigma_\alpha \equiv (\omega/k_\parallel)[m_\alpha/(2T_{\alpha 0})]^{1/2}$, and introduced $W(\varsigma_\alpha) \equiv [dZ(\varsigma_\alpha)/d\varsigma_\alpha]/2 \equiv -1-\varsigma_\alpha Z(\varsigma_\alpha)$, $Z(\varsigma_\alpha)$ being the plasma dispersion function[4].

**References**

1. B. Basu, *Theory of Collective Excitations in Plasma* (Amazon Publishing, 2025), p. 76.

2. P. J. Catto, A. M. El-Nadi, C. S. Liu, and M. N. Rosenbluth, *Nucl. Fusion* 14, 405 (1974).

3. N. A. Krall and A. W. Trivelpiece, *Principles of Plasma Physics* (McGraw-Hill, New York, 1973), p. 418.

4. B. D. Fried and S. D. Conte, *The Plasma Dispersion Function* (Academic, New York, 1961).